\documentclass[prl,superscriptaddress,preprint,endfloats,nopacs]{revtex4}%preprint type
\usepackage[dvipdfmx]{graphicx}
\usepackage{bm}
\usepackage{pifont}
\usepackage{amsmath, amsthm, amssymb}
\newcommand {\ECS}{EuCd$_2$Sb$_{\mathrm{2}}$}
\newcommand {\ECA}{EuCd$_2$As$_{\mathrm{2}}$}
\newcommand {\CSS}{Co$_{3}$Sn$_2$S$_2$}
\newcommand {\MS}{Mn$_{3}$Sn}

\newcommand {\MG}{Mn$_3$Ge}

\newcommand {\etal}{\textit{et al}.}

\newcommand {\EF}{$E_{\mathrm{F}}$}
\newcommand {\rhoxx}{$\rho_\mathrm{xx}$}

\newcommand {\TN}{$T_{\mathrm{N}}$}

\begin{document}
	\title{
		Maximizing intrinsic anomalous Hall effect by controlling\\
		the Fermi level in simple Weyl semimetal films
	}
	%\author{M. Ohno}
	\author{Mizuki Ohno}
	\affiliation{Department of Applied Physics, University of Tokyo, Tokyo 113-8656, Japan}
	\affiliation{Quantum-Phase Electronics Center (QPEC), University of Tokyo, Tokyo 113-8656, Japan}
	\affiliation{Department of Physics, Tokyo Institute of Technology, Tokyo 152-8550, Japan}
	%\author{S. Minami}
	\author{Susumu Minami}
	\affiliation{Department of Physics, University of Tokyo, Tokyo 113-8656, Japan}
	%\author{Y. Nakazawa}
	\author{Yusuke Nakazawa}
	\affiliation{Department of Applied Physics, University of Tokyo, Tokyo 113-8656, Japan}
	\affiliation{Quantum-Phase Electronics Center (QPEC), University of Tokyo, Tokyo 113-8656, Japan}
	%\author{S. Sato}
	\author{Shin Sato}
	\affiliation{Department of Applied Physics, University of Tokyo, Tokyo 113-8656, Japan}
	\affiliation{Quantum-Phase Electronics Center (QPEC), University of Tokyo, Tokyo 113-8656, Japan}
	%\author{M. Kriener}
	\author{Markus Kriener}
	\affiliation{RIKEN Center for Emergent Matter Science (CEMS), Wako 351-0198, Japan}
	%\author{R. Arita}
	\author{Ryotaro Arita}
	\affiliation{Department of Applied Physics, University of Tokyo, Tokyo 113-8656, Japan}
	\affiliation{RIKEN Center for Emergent Matter Science (CEMS), Wako 351-0198, Japan}
	%\author{M. Kawasaki}
	\author{Masashi Kawasaki}
	\affiliation{Department of Applied Physics, University of Tokyo, Tokyo 113-8656, Japan}
	\affiliation{Quantum-Phase Electronics Center (QPEC), University of Tokyo, Tokyo 113-8656, Japan}
	\affiliation{RIKEN Center for Emergent Matter Science (CEMS), Wako 351-0198, Japan}
	%\author{M. Uchida}
	\author{Masaki Uchida}
	\email[Author to whom correspondence should be addressed: ]{m.uchida@phys.titech.ac.jp}
	\affiliation{Department of Physics, Tokyo Institute of Technology, Tokyo 152-8550, Japan}
	\affiliation{PRESTO, Japan Science and Technology Agency (JST), Tokyo 102-0076, Japan}
	\begin{abstract}
		Large intrinsic anomalous Hall effect (AHE) originating in the Berry curvature has attracted growing attention for potential applications.
		Recently proposed magnetic Weyl semimetal {\ECS} provides an excellent platform for controlling the intrinsic AHE because it only hosts a Weyl-points related band structure near the Fermi energy.
		Here we report the fabrication of {\ECS} single-crystalline films and control of their anomalous Hall effect by film technique.
		As also analyzed by first-principles calculations of energy-dependent intrinsic anomalous Hall conductivity, the obtained anomalous Hall effect shows a sharp peak as a function of carrier density, demonstrating clear energy dependence of the intrinsic AHE.
	\end{abstract}
	\maketitle
	Topological semimetals, characterized by massless Dirac band dispersion in the low energy excitation, exhibit unique magnetotransport due to their nontrivial band topology in the momentum space \cite{Burkov2011,Wang2017f,Armitage2018,Lv2021}.
	Recently reported magnetic topological semimetals further enrich the understanding of magnetotransport, where entanglement between magnetism and nontrivial topology can lead to exotic magnetotransport such as a large anomalous Hall effect (AHE) \cite{Miyasato2007,Nagaosa2010,Xiao2010}.
	Integration of the Berry curvature for all occupied states gives rise to a large intrinsic contribution to anomalous Hall conductivity.
	Namely, anomalous Hall conductivity is closely related to the electron occupation determined by the position of Fermi level \cite{Thouless1982, Xiao2010}.
	
	Paired Weyl points (WPs) in magnetic Weyl semimetals can be viewed as monopoles with opposite chirality corresponding to the sources (W$+$) and sinks (W$-$) of the Berry curvature (Fig.~1(a)).
	Due to the Berry flux of paired WPs, anomalous Hall conductivity $\sigma_\mathrm{xy, AHE}$ is expected to  exhibit  a large peak at the energy position of WPs $(E_\mathrm{W})$ \cite{Burkov2011a,Lu2015} and be strongly dependent on the Fermi level.
	Specifically, $\sigma_\mathrm{xy, AHE}$ can be maximized if the Fermi level can be adjusted to $E_\mathrm{W}$~\cite{Noky2018,Muechler2020}.
	However, such an energy dependence of $\sigma_\mathrm{xy, AHE}$ has not been demonstrated yet due to high carrier densities ($\sim 10^{20}$--$10^{22}$~cm$^{-3}$) and complicated band structures of conventional magnetic Weyl semimetals, while the modulation of $\sigma_\mathrm{xy, AHE}$ has been attempted by chemical doping and pressure application for {\CSS} \cite{Shen2020,Shen2020b,Zhou2020a,Thakur2020}, {\MS} \cite{Kuroda2017a,Khadka2020,Chen2021e,Chen2021g}, and {\MG} \cite{Kiyohara2016}.
	
	In this context, {\ECA} and isostructural {\ECS}, where one of three Cd atoms is substituted with Eu in the parent Dirac semimetal Cd$_3$As$_2$, provide an ideal platform for experimentally demonstrating the peculiar energy dependence of intrinsic anomalous Hall conductivity.
	This is because they have a simple band structure near the Fermi energy (\EF) and much lower carrier densities ($\sim 10^{18}$--$10^{19}$ cm$^{-3}$) \cite{Soh2019,Su2020}.
	On the other hand, difficulties in the growth of thin films have prevented the study of carrier density dependence with taking these advantages.
	
	\begin{figure}
		\begin{center}
			\includegraphics*[bb=0 0 336 434,width=11cm]{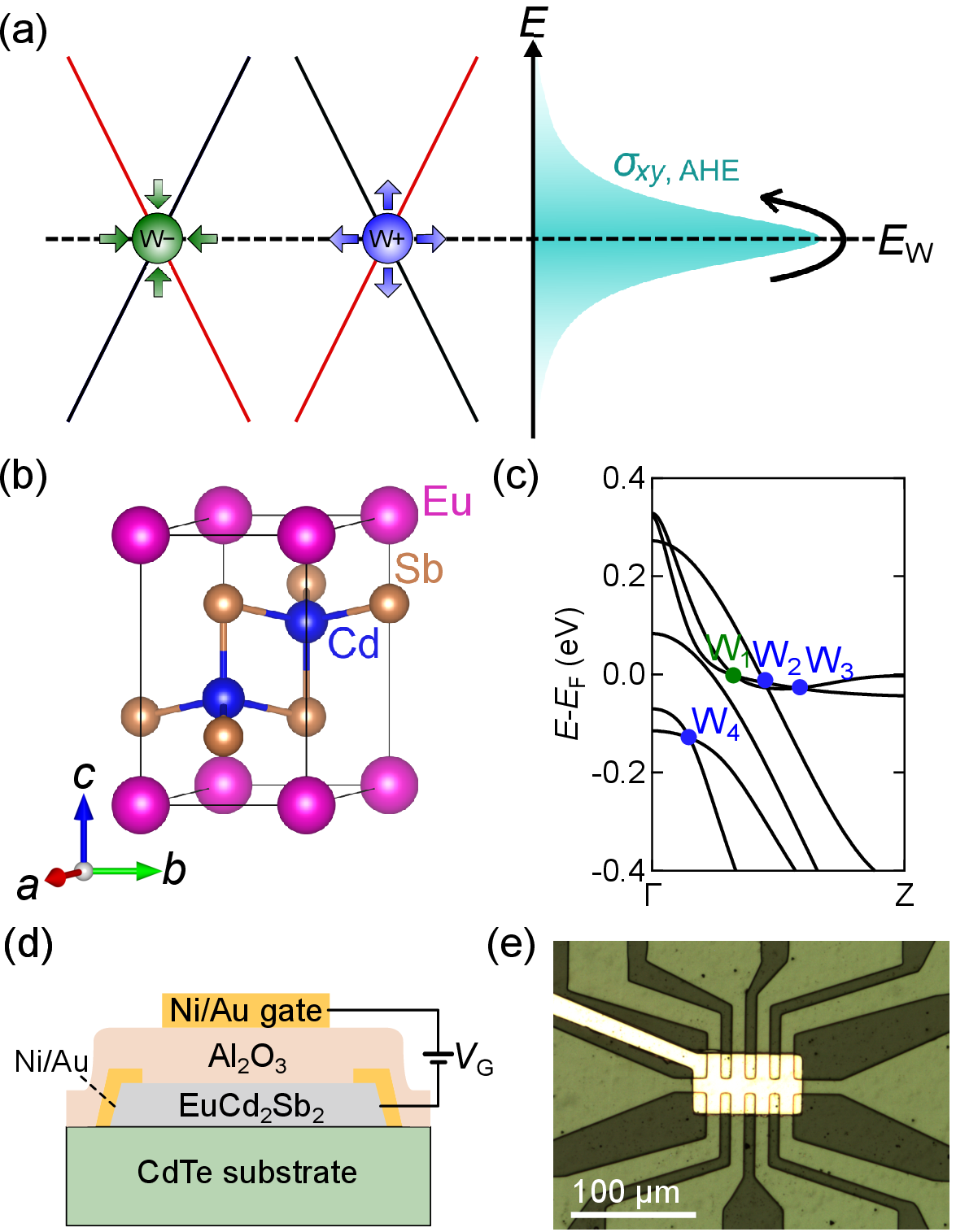}
			\caption{
				(a) Energy dispersion of paired Weyl points (WPs) and energy-dependent anomalous Hall conductivity $\sigma_{xy,{\mathrm{ AHE}}}$. 
				Weyl points W$+$/W$-$ are the source/sink of the Berry curvature.
				(b) Crystal structure of {\ECS} \cite{Artmann1996} drawn using VESTA \cite{Momma2011}.
				(c) Energy dispersion calculated for {\ECS} films, showing four pairs of WPs near the Fermi energy.
				They emerge along the $\Gamma-$Z direction in the Brillouin zone when magnetic moments are forcedly aligned along the $c$ axis.
				Blue/green circles represent WPs (W$_1$-W$_4$) with plus/minus chirality.
				(d) Cross-sectional schematic illustration and (e) top-view photograph of the electrostatic gate device fabricated from a  {\ECS} film, where holes are depleted (accumulated) when $V_\mathrm{G}$ is positively (negatively) biased.
			}
			\label{fig1}
		\end{center}
	\end{figure}
	
	{\ECS} has a trigonal crystal structure with space group $P$-$3m1$ \cite{Artmann1996,Goryunov2012,Soh2018b,Su2020}.
	It consists of an alternate stacking of triangular Eu layers and Cd$_2$Sb$_2$ layers (Fig.~1(b)).
	Eu atoms contribute to magnetism in this compound, forming $A$-type antiferromagnetic structure below the N\'{e}el temperature $T_\mathrm{N}=7.4$~K with Eu$^{2+}$ magnetic moments lying in the $ab$ plane \cite{Goryunov2012, Soh2018b}.
	As shown in Fig.~1(c), four pairs of WPs (W$_1$-W$_4$) near {\EF} along the $k_z$ direction are protected by $C_{3z}$ symmetry when the magnetic Eu moments are ferromagnetically polarized along the $c$ axis under the out-of-plane magnetic field \cite{Su2020}.
	There are no other bands near {\EF} and all the hole carriers occupy the WPs-related valence bands.
	Therefore, it is expected that the Fermi level dependence of intrinsic anomalous Hall conductivity can be demonstrated by modulating hole carrier densities in {\ECS} with film techniques.
	
	Here we report the fabrication of single-crystalline {\ECS} films by molecular beam epitaxy and investigation of their anomalous Hall effect using film techniques.
	By adjusting the Fermi level with controlling Sb flux conditions and also performing electrostatic gating experiments, we find that their anomalous Hall effect is largely enhanced compared to bulks and strongly dependent on the carrier density with a sharp peak, as also demonstrated by first-principles calculations.
	
	{\ECS} films were grown in an Epiquest RC1100 chamber \cite{Nakazawa2019,Ohno2021,Ohno2021a,Uchida2021} on single-crystalline (111)A CdTe substrates.
	In-plane lattice mismatch between {\ECS} and CdTe is 2.4\% as shown in Fig.~2(b).
	CdTe substrates were etched with 0.01\% Br$_{2}$-methanol before loading it into the chamber and then heated to 750$^{\circ}$C with Cd flux supplied to obtain an atomically smooth surface \cite{Nakazawa2019}.
	After annealing the substrate, it was cooled to the growth temperature of 360$^{\circ}$C.
	The beam equivalent pressures were measured by an ionization gauge, and were set to $1.2{\times}10{^{-5}}$~Pa for Eu, 5.0$\times10^{-4}$~Pa for Cd, and $8.5\times10^{-6}$~Pa for Sb during the co-deposition, which is the same growth condition as EuCdSb$_2$ films on Al$_2$O$_3$ (0001) substrates \cite{Ohno2021a}.
	This Cd-rich growth condition offers advantages in reducing carrier densities in {\ECS} films, because Cd deficiency is a major origin of the hole carriers.
	The films were then annealed $in$-$situ$ at 500$^{\circ}$C for 5 minutes under exposure of Cd flux.
	The film thickness was set at 50~nm, and the growth rate was about 0.07~{\AA}/s.
	
	The films were patterned to Hall bars with a channel width of 10 $\mu$m through conventional photolithography, Ar ion milling, and chemical wet etching processes: 10~nm thick Ni and 50~nm thick Au electrodes were deposited for ohmic contact by electron beam evaporation, and then 30~nm thick Al$_2$O$_3$ was deposited as gate dielectric by atomic layer deposition, as shown in Figs.~1(d) and 1(e).
	Low-temperature magnetotransport was measured using a Quantum Design Physical Property Measurement System cryostat equipped with a 9~T superconducting magnet.
	Hall measurements were performed using a lock-in technique.
	The excitation current was kept constant at 0.5~mA with a frequency of 13~Hz.
	For gating experiments, DC bias ($V_\mathrm{G}=-8 \sim +8$~V) voltage was applied to the gate electrode.
	
	First-principles calculations of the band structure with spin-orbit coupling were performed using the VASP package \cite{Kresse1996, Kresse1996a, Kresse1999} for experimentally determined film lattice parameters.
	The generalized gradient approximation of Perdew-Burke-Ernzerhof was adopted for the exchange-correlation functional \cite{Perdew1996}.
	The correlation effect was considered by a Hubbard $U$ correction of $U = 4.5$~eV for the Eu-$f$ orbitals \cite{Dudarev1998}, which was determined by comparing the band structures revealed by angle-resolved photoemission spectroscopy and first-principles calculations \cite{Su2020}.
	$16\times16\times10$~$k$-point mesh with Monkhorst-Pack scheme \cite{Monkhorst1976} was used for the Brillouin zone sampling of the primitive cell and Gaussian smearing with a width of $0.02$~eV was applied.
	From the Bloch states obtained in the DFT calculation, a Wannier basis set is constructed by using the Wannier90 code \cite{Pizzi2020}.
	The basis consists of $s$, $f$-character orbitals localized at the Eu site, $s$, $d$-character orbitals at the Cd, and $s$, $p$-character orbitals at the Sb.
	The intrinsic anomalous Hall conductivity is computed using a $k$-point mesh of $300\times300\times300$ \cite{Wu2018}.
		
	\begin{figure}
		\begin{center}
			\includegraphics*[bb=0 0 723 348,width=17cm]{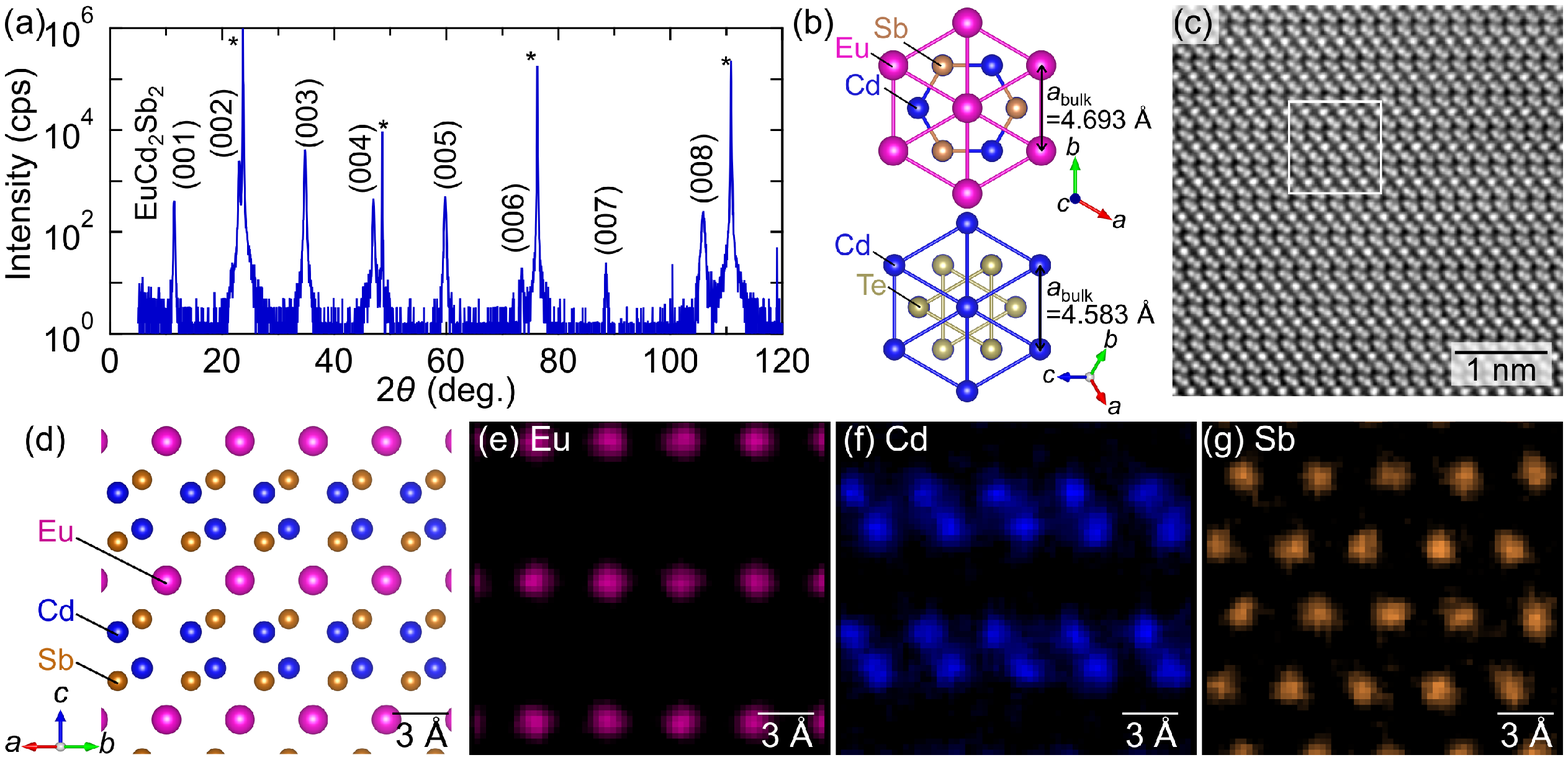}
			\caption{
				(a) XRD $\theta$-2$\theta$ scan of a {\ECS} film grown on a CdTe substrate.
				Substrate peaks are marked with an asterisk.
				(b) Crystal structures of {\ECS} film and CdTe substrate, viewed along the out-of-plane direction.
				(c) Cross-sectional image of {\ECS} film, taken by high-angle annular dark-field scanning transmission electron microscopy.
				(d) Atomic arrangement of {\ECS} viewed along [110] direction.
				Energy dispersive x-ray spectrometry maps taken for (e) Eu $L$, (f) Cd $L$, and (g) Sb $L$ edge in the boxed region in (c).
			}
			\label{fig2}
		\end{center}
	\end{figure}
	
	Figure~2 shows structural characterization of {\ECS} film.
	As shown in the x-ray diffraction (XRD) $\theta$-2$\theta$ scan in Fig.~2(a), reflections from the (001) {\ECS} lattice planes are observed without any impurity phases.
	Its out-of-plane lattice constant along the $c$-axis is calculated to be $7.73$~{\AA}.
	Figure~2(c) shows a cross-section image of the {\ECS}, taken by high-angle annular dark-field scanning transmission electron microscopy.
	The periodic atomic arrangement corresponding to the {\ECS} crystal structure in Fig.~2(d) is clearly confirmed.
	Elemental maps are also taken by energy dispersive x-ray spectrometry, as shown in Figs.~2(e)-2(g).
	The alternate stacking of Eu layers and Cd$_2$Sb$_2$ layers is clearly resolved.
	Further structural characterization was performed for examining the in-plane epitaxial relation (See Supplementary Materials for details \cite{SM}).
	
	\begin{figure}
	\newpage
		\begin{center}
		 	\includegraphics*[bb=0 0 474 558,width=12cm]{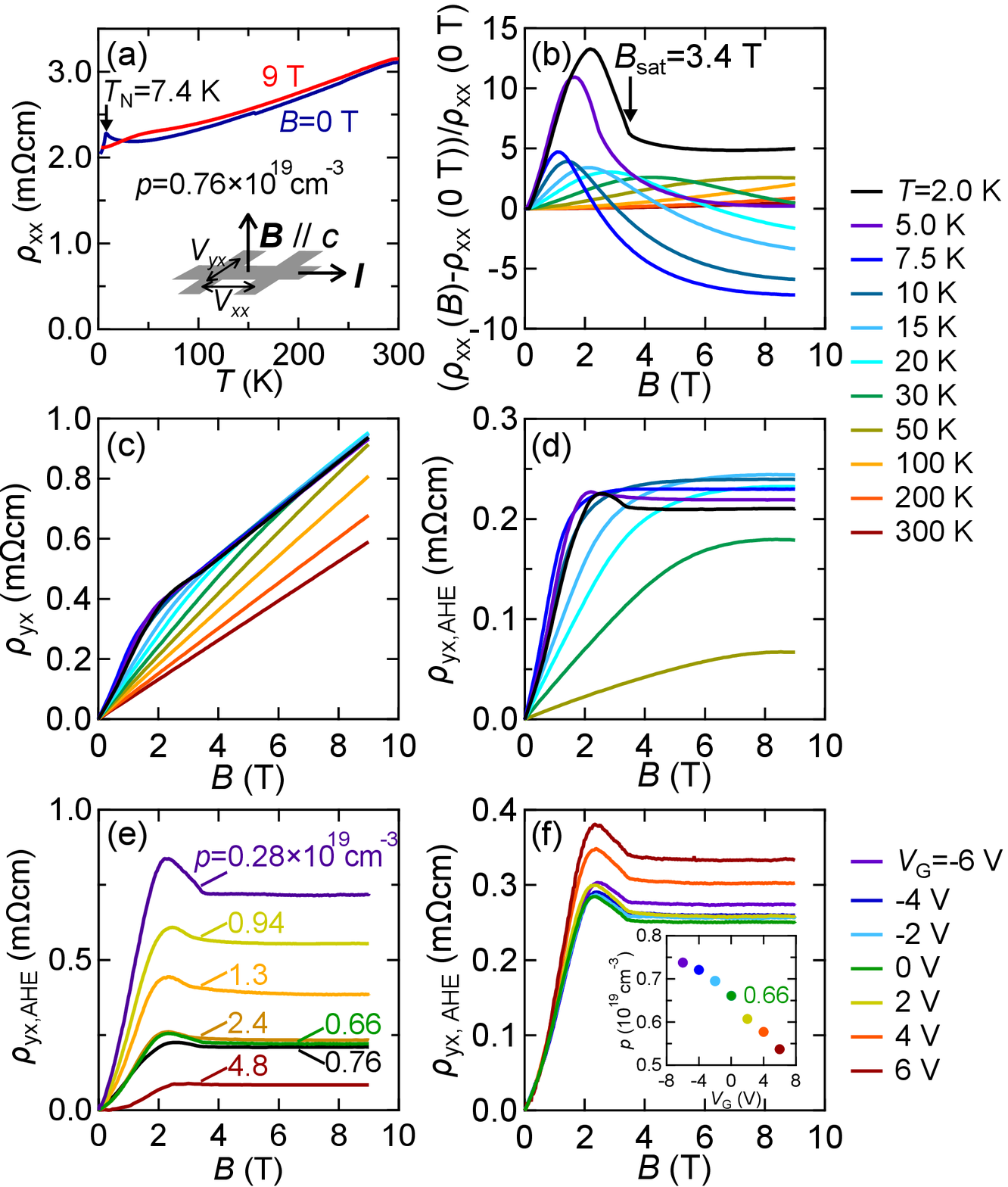}
			\caption{
				Magnetotransport measured for a {\ECS} film with a carrier density $p=0.8\times10^{19}$~cm$^{-3}$.
				(a) Temperature dependence of longitudinal resistivity $\rho_\mathrm{xx}$, with a clear kink at the N\'{e}el temperature $T_\mathrm{N}=7.4$~K.
				The inset shows measurement configuration.
				(b) Magnetoresistance $(\rho_\mathrm{xx}(B)-\rho_\mathrm{xx}(0$~$\mathrm{T}))/\rho_\mathrm{xx}(0$~$\mathrm{T})$ taken with sweeping the out-of-plane magnetic field at various temperatures.
				The saturation field is estimated $B_\mathrm{sat}=3.4$~T at $T=2.0$~K.
				(c) Hall resistivity $\rho_\mathrm{yx}$ taken at various temperatures.
				(d) Anomalous Hall resistivity $\rho_\mathrm{yx, AHE}$, obtained by subtracting the ordinary term at various temperatures.
				Anomalous Hall resistivity $\rho_\mathrm{yx, AHE}$ of {\ECS} films with (e) different carrier densities and (f) electrostatic gating at 2.0 K. 
			}
			\label{fig3}
		\end{center}
	\end{figure}
	
	Figure~3 summarizes fundamental magnetotransport of the {\ECS} film.
	As shown in Fig.~3(a), longitudinal resistivity {\rhoxx} exhibits a metallic behavior in the whole temperature regime.
	Below 50~K, {\rhoxx} gradually increases upon cooling until it shows a clear kink at 7.4~K.
	Then it exhibits a sudden drop upon further cooling.
	This kink observed at 7.4~K corresponds to N\'{e}el temperature {\TN} \cite{Goryunov2012, Soh2018b,Su2020}.
	The kink disappears at the out-of-plane magnetic field $B=9$~T, where the Eu$^{2+}$ magnetic moments are completely aligned along the $c$-axis \cite{Soh2018b,Su2020}.
	This low temperature behavior can be understood considering the scattering of conduction electrons by the localized Eu$^{2+}$ spins.
	Figure~3(b) presents magnetoresistance (MR) taken by sweeping the out-of-plane magnetic field at various temperatures.
	At $T = 2.0$~K, a peak ends at the saturation field of $B_\mathrm{sat}=3.4$~T, and which shifts to lower fields upon increasing temperature and then disappears above ${T_\mathrm{N}}$.
	This peak below $B_\mathrm{sat}$ is ascribed to an increase of magnetic scattering in the canted spin structure.
	At high-fields above $B_\mathrm{sat}$, the Eu$^{2+}$ spins are fully polarized along the $c$-axis, resulting in reduced magnetic scattering.
	At higher temperatures above \TN, the peak becomes smaller and shifts to higher magnetic fields.
	Finally, the peak disappears and conventional positive MR appears.
	
	Figure~3(c) shows Hall resistivity $\rho_\mathrm{yx}$ measured at various temperatures.
	At temperatures above 50~K, Hall resistivity depends almost linearly on the magnetic field.
	In contrast, it deviates from the linear behavior upon cooling below 50~K.
	Here $\rho_\mathrm{yx}$ is expressed as $\rho_\mathrm{yx}=R_\mathrm{H} B+\rho_\mathrm{yx, AHE}$ with the ordinary Hall coefficient $R_{\mathrm{H}}$ and the anomalous Hall resistivity $\rho_\mathrm{yx, AHE}$.
	Deviation from the linear behavior disappears above $B_\mathrm{sat}$, showing clear contribution of the anomalous Hall term (See Supplementary Materials [39]).
	By subtracting the ordinary term with a linear fit, the anomalous term is obtained as shown in Fig.~3(d).
	$\rho_\mathrm{yx, AHE}$ taken at $T = 2.0$~K exhibits a kink at $B_\mathrm{sat}=3.4$~T.
	With increasing temperature, $B_\mathrm{sat}$ shifts to lower fields below {\TN} and then disappears above {\TN}, as confirmed in MR.
	Besides, a broad but pronounced peak appears around $B_\mathrm{sat}$ below $T_\mathrm{N}$, showing deviation from the magnetization behavior.
	While this peak suggests the existence of the so-called topological Hall term, it is unlikely that this is caused by the real-space spin Berry phase on noncoplanar spin configuration during simple spin polarization process.
	Rather, this may be understood by intrinsic AHE corresponding to band structure changes during the polarization process, as also reported for other Eu compounds \cite{Cao2021,Takahashi2018,Mayo2021,Shen2021}.
	
	Here we define the saturation value of $\rho_\mathrm{yx, AHE}$ at 9~T as $\rho_{\mathrm{yx, AHE}}^\mathrm{sat}$.
	It is expected that $\rho_\mathrm{yx, AHE}$ is dominated by intrinsic AHE in the forced ferromagnetic phase.
	In addition, $\sigma_\mathrm{xx}$ and $\sigma_\mathrm{xy, AHE}$ are calculated by
	\begin{align}
	\sigma_\mathrm{xx}&=\frac{\rho_\mathrm{xx}(B=0\mathrm{~T})}{\left(\rho_{\mathrm{yx, AHE}}^\mathrm{sat}\right)^{2}+\rho_\mathrm{xx}(B=0\mathrm{~T})^{2}},
	\end{align}
	and
	\begin{align}
	\sigma_\mathrm{xy, AHE}&=\frac{\rho_{\mathrm{yx, AHE}}^\mathrm{sat}}{\left(\rho_{\mathrm{yx, AHE}}^\mathrm{sat}\right)^{2}+\rho_\mathrm{xx}(B=0\mathrm{~T})^{2}}.
	\end{align}
	AHE has been roughly categorized into three regimes based on the range of $\sigma_\mathrm{xx}$~\cite{Nagaosa2010, Onoda2006}.
	For {\ECS} films in this study, $\sigma_\mathrm{xx}$ is below $1 \times 10^4$~$\Omega^{-1}$cm$^{-1}$, and the carrier density is located between $0.3 \times 10^{19}$ and $4.8 \times 10^{19}$~cm$^{-3}$.
	This suggests that the anomalous Hall velocity induced by the Berry curvature is suppressed with following the scaling relation $\sigma_\mathrm{xy, AHE} \propto \sigma_\mathrm{xx}^{1.6}$.
	With reducing the hole carrier density of {\ECS} films at 2.0~K, $\rho_\mathrm{yx, AHE}$  increases at first, decreases, and then increases again as shown in Fig.~3(e).
	As confirmed in Fig.~3(f), modulating carriers with electrostatic gating also follows this trend. 
			
	\begin{figure}
		\begin{center}
			\includegraphics*[bb=0 0 417 608,width=11cm]{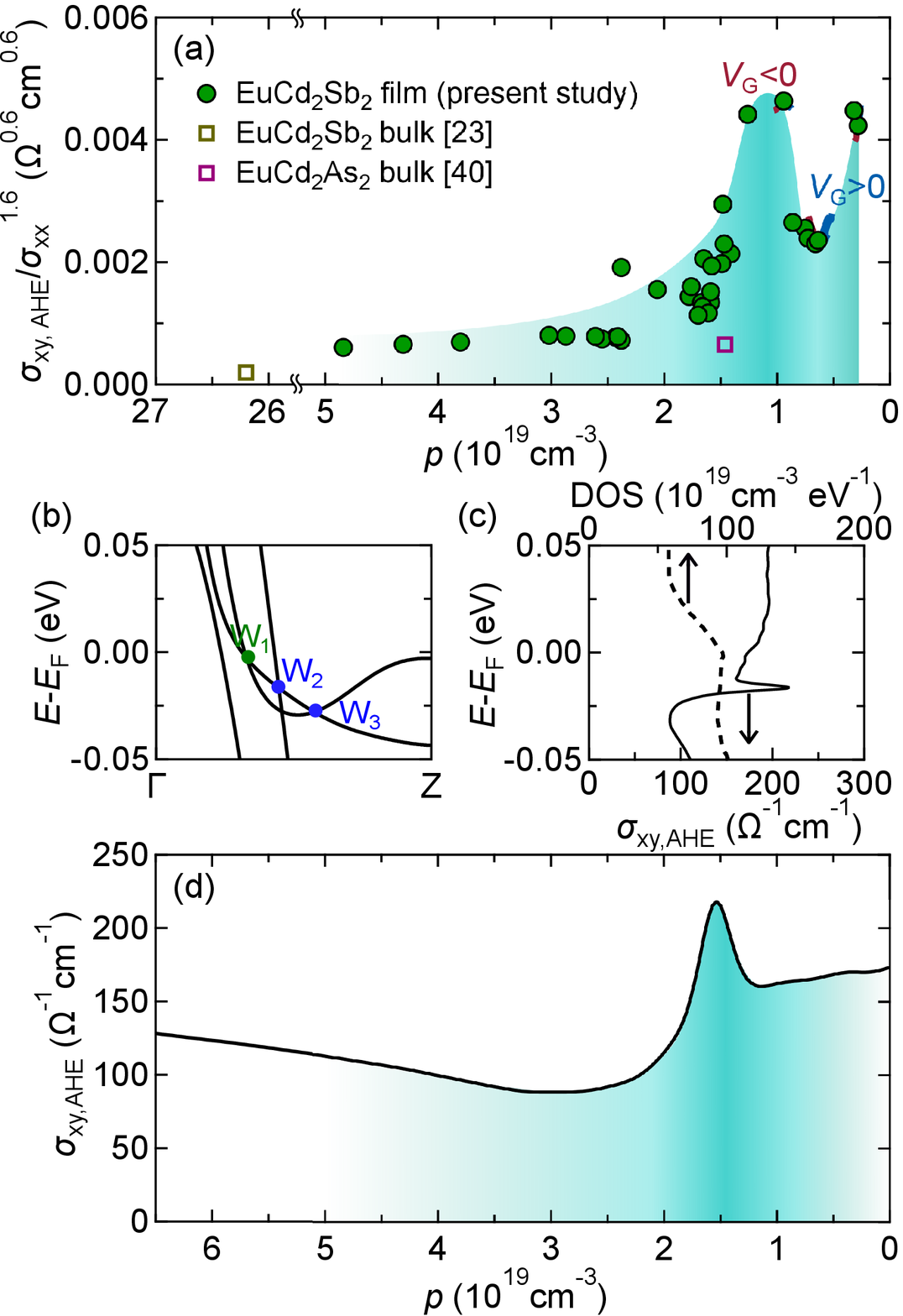}
			\caption{
				(a) Carrier density dependence of $\sigma_\mathrm{xy, AHE}/\sigma_{xx}^{1.6}$ in {\ECS} films, compared to bulks of  {\ECS} \cite{Su2020} and {\ECA} \cite{Cao2021}.
				Electrostatic gating data obtained for low carrier samples ($p < 1 \times 10^{19}$~cm$^{-3}$) are also plotted with curves, colored by the sign of $V_\mathrm{G}$.
				(b) Magnified energy dispersion.
				(c) $\sigma_\mathrm{xy, AHE}$ and density of states (DOS) calculated as a function of energy. 
				(d) $\sigma_\mathrm{xy, AHE}$ plotted for carrier density converted from DOS.
			}
			\label{fig4}
		\end{center}
	\end{figure}
	
	In order to discuss the carrier density dependence of AHE in the dirty metal regime \cite{Nagaosa2010, Onoda2006}, we show the carrier density dependence of $\sigma_\mathrm{xy, AHE}/\sigma_{xx}^{1.6}$ obtained by measuring {\ECS} films at 2.0 K, as shown in Fig.~4(a).
	Remarkably, $\sigma_\mathrm{xy, AHE}/\sigma_{xx}^{1.6}$ exhibits a large peak at about $p = 1 \times 10^{19}$~cm$^{-3}$ and the same trend is confirmed by continuously modulating the carrier density with electrostatic gating.
	We also confirm that a similar large peak is observed in the plots of $\sigma_\mathrm{xy, AHE}$ and $\sigma_\mathrm{xy, AHE}/\sigma_{xx}$ (See Supplementary Materials [39]).
	$\sigma_\mathrm{xy, AHE}/\sigma_{xx}^{1.6}$ values previously reported for bulk single crystals \cite{Su2020} also follow the same trend as revealed by the set of film measurements.
	Therefore, the observed peak is ascribed to successful adjustment of the Fermi level to $E_\mathrm{W}$.
	
	In order to understand the enhancement of $\sigma_\mathrm{xy, AHE}/\sigma_{xx}^{1.6}$ more quantitatively, we perform first-principles calculations of band structure and intrinsic $\sigma_\mathrm{xy, AHE}$.
	As shown in Fig.~4(c), $\sigma_\mathrm{xy, AHE}$ exhibits a sharp peak at about $E-E_\mathrm{F}=-0.02 $~eV, which corresponds to the energy position of W$_2$ and W$_3$ in Fig.~4(b).
	The energy can be converted to carrier density by integrating density of states from $E-E_\mathrm{F}=0$~eV.
	As confirmed in Fig.~4(d), $\sigma_\mathrm{xy, AHE}$ exhibits clear carrier density dependence consistent with the experimental data, demonstrating that the Fermi level of {\ECS} films passes through the energy position of W$_2$ and W$_3$.
		
	In summary, we have systematically investigated carrier density dependence of the intrinsic anomalous Hall conductivity by adjusting the Fermi level in simple magnetic Weyl semimetal {\ECS} films.
	The peak observed for $\sigma_\mathrm{xy, AHE}/\sigma_{xx}^{1.6}$  has been also reproduced by first-principles calculations of energy-dependent intrinsic anomalous Hall conductivity.
	Our findings have provided transport evidence of intrinsic anomalous Hall conductivity originating in WPs.
	The present work paves the way for further exploring the potential of WPs-based exotic magnetotransport using film techniques.
	
	This work was supported by JST PRESTO Grant No. JPMJPR18L2 and JST CREST Grant No. JPMJCR16F1, Japan and by Grant-in-Aid for Scientific Research (B) No. JP21H01804 from MEXT, Japan, and by TEPCO Memorial Foundation, Japan.
	The data that support the findings of this study are available from the corresponding author upon reasonable request.

\end{document}